\documentclass[twocolumn,amssymb,amsmath,superscriptaddress,aps,pre,reprint]{revtex4-1}
\usepackage{graphicx}
\usepackage{dcolumn}
\usepackage{bm}
\usepackage{subfigure}

\begin {document}

\title{Mean-field dynamics of a population of stochastic map neurons}

\author{Igor Franovi\'c}
\email{franovic@ipb.ac.rs}
\affiliation{Scientific Computing Laboratory, Center for the Study of Complex Systems,
Institute of Physics Belgrade, University of Belgrade, Pregrevica 118, 11080 Belgrade, Serbia}

\author{Oleg V. Maslennikov}
\email{olmaov@ipfran.ru}
\affiliation{Institute of Applied Physics of the Russian Academy of Sciences, 46 Ulyanov Street,
603950 Nizhny Novgorod, Russia}

\author{Iva Ba\v{c}i\'c}
\affiliation{Scientific Computing Laboratory, Center for the Study of Complex Systems,
Institute of Physics Belgrade, University of Belgrade, Pregrevica 118, 11080 Belgrade, Serbia}

\author{Vladimir I. Nekorkin}
\affiliation{Institute of Applied Physics of the Russian Academy of Sciences,
46 Ulyanov Street, 603950 Nizhny Novgorod, Russia}

\date{\today}

\begin{abstract}
We analyze the emergent regimes and the stimulus-response relationship of a population of noisy map neurons by means of a mean-field model, derived within the framework of cumulant approach complemented by the Gaussian closure hypothesis. It is demonstrated that the mean-field model can qualitatively account for stability and bifurcations of the exact system, capturing all the generic forms of collective behavior, including macroscopic excitability, subthreshold oscillations, periodic or chaotic
spiking and chaotic bursting dynamics. Apart from qualitative analogies, we find a substantial quantitative agreement between the exact and the approximate system, as reflected in matching of the parameter domains admitting the different dynamical regimes, as well as the characteristic properties of the associated time series. The effective model is further shown to reproduce with sufficient accuracy the phase response curves of the exact system and the assembly's response to external stimulation of finite amplitude and duration.
\end{abstract}

\pacs{05.45.Xt, 89.75.Fb, 05.40.Ca}

\maketitle

Cortical connectivity patterns exhibit a hierarchical modular organization from the microscopic level of
interacting neurons, via cortical columns and other types of mesoscopic circuitry, up to fibers projecting between the distributed brain areas \cite{ZZHK06,BS09,MLB10}. Architecture of neural assemblies comprising the anatomical modules closely reflects the functional specialization over different modalities \cite{SCKH04}, such that the macroscopic dynamics of neuronal populations, as well as the interplay of the associated collective modes, underpin various stages of information processing and higher cognitive functions \cite{MPS10}. In terms of evoked activity in the cortex, an ample example concerns the sensory regions, where many neurons display similar responses to a given stimulus, which strongly indicates that the information content is primarily encoded by the collective assembly response \cite{ALP06}. With regard to self-organized dynamics, the different forms of synchronization between spiking or bursting neurons give rise to macroscopic oscillations known to span several orders of magnitude in frequency range \cite{VLRM01,B09}. Such oscillations are deemed to facilitate coordinated rhythmic tasks and provide the dynamical background behind perception, attention, memory consolidation, motor planning or sleep \cite{VLRM01,B09,VW09}. Nonetheless, interaction between multiple rhythms has been indicated as critical for merging the activities of distant brain regions through cross-frequency coupling and other mechanisms \cite{CS06,LJ13}. For these reasons, gaining a deeper understanding of macroscopic behavior of neural populations has become an outstanding issue in neuroscience, focused primarily on the mechanisms guiding the emergent collective dynamics and the fashion in which assemblies respond to various external stimuli. Conceptually, such an approach is reminiscent of a common paradigm in nonlinear dynamics, where systems of coupled oscillators exhibiting a collective mode are typically treated as macroscopic oscillators, which may then be subjected to an external drive or can be influenced by collective rhythms from afferent populations \cite{BRZK09}.

In the present paper, we systematically analyze the emergent dynamics and the stimulus-response relationship of a population of stochastic map neurons using a mean-field ($MF$) approach. The considered map neurons can exhibit a variety of regimes, including excitability, subthreshold oscillations, regular and chaotic spiking or bursting, as well as mixed spiking-bursting oscillations \cite{MN14,MKRN13,MNK15,MN15}. Despite that the collective motion of spiking or bursting neurons subjected to noise has been extensively studied using different models of discrete local dynamics, such as Rulkov \cite{R02,RTB04,WL07,WDPC08,BPVL07,ICS11,FM10,FM11} or Izhikevich neuron maps \cite{I06,IE08}, a $MF$ theory for a population of stochastic map neurons is obtained here for the first time. Nevertheless, in case of continuous time systems, the $MF$ approach has been a standard analytical tool for treating diverse problems in neuroscience and other fields \cite{FB05,LTGE02,B10,BC07,BH99,H03}. Note that our derivation of the effective model relies on Gaussian approximation, which is introduced within the framework of a Gaussian closure hypothesis \cite{LGNS04,FTVB14,FTVB13,FTVB12,KF15,ZSSN05,SZNS13}.

The particular set of issues we address consists in establishing whether and how the $MF$ model can be used to $(i)$ qualitatively analyze the network stability and bifurcations of the exact system associated to emergence of generic macroscopic regimes; $(ii)$ provide adequate quantitative predictions in terms of bifurcation thresholds, and the average interspike intervals or bursting cycles of the exact system, as well as $(iii)$ accurately anticipate the population's response to different forms of external stimuli. Within this context, it will be examined whether the effective model is capable of reproducing the properties of noise-activated, noise-induced and noise-perturbed modes of collective behavior.

The paper is organized as follows. In Section \ref{sec:Model}, we make an overview of the local map dynamics and introduce the population model. Section \ref{sec:MF} outlines the ingredients most relevant for the derivation of the $MF$ system, with the remaining technical details left for the Appendix. In Section \ref{sec:Standbif}, the qualitative and quantitative agreement between the dynamics of the exact and the $MF$ model is illustrated by the appropriate bifurcation diagrams, as well as by comparing the characteristic features of the associated regimes. Section \ref{sec:Responses} concerns the assembly's stimulus-response relationship, first investigating the analogy between the respective phase-response curves ($PRCs$) of the exact system and the effective model in spiking and bursting regimes, and then considering the extent to which the $MF$ model reproduces the population's response to rectangular pulses of finite amplitude and duration. In Section \ref{sec:Summary}, we provide a summary of our main results.

\section{Map neuron dynamics and the population model} \label{sec:Model}

The dynamics of an isolated neuron conforms to a map model first introduced in \cite{NV07,CNV07}, which is given by
\begin{align} \label{eq1}
x_{n+1}&=x_{n}+G(x_{n})-\beta H(x_{n}-d)-y_{n}, \\ \nonumber
y_{n+1}&=y_{n}+\epsilon (x_{n}-J),
\end{align}
where $n$ denotes the iteration step. The variable $x_n$ qualitatively accounts for the membrane potential, whereas the recovery variable $y_n$, whose rate of change is set by a small parameter $\epsilon=10^{-2}$, mimics the behavior of ion-gating channels. The parameters $a,\beta$ and $d$
modify the profile of the ensuing oscillations, while $J$ crucially influences the neural
excitability, viz. the transitions from silence to active regimes.

The $x_n$ evolution features two nonlinear terms, one being a FitzHugh-Nagumo-like cubic nonlinearity $G(x_n)=x_n(x_n-a)(1-x_n)$, which is complemented by a discontinuity term $-\beta H(x_{n}-d)$, where $H$ stands for the Heaviside step function. The parameters $a=0.1$ and $d=0.45$ are kept fixed throughout the paper. The impact of discontinuity consists in making the fast subsystem (Eq. \eqref{eq1} with $\epsilon=0$) a Lorenz-type map within certain parameter domains \cite{CNV07,MN13}, which endows the model with the ability to generate chaotic spike or burst oscillations, otherwise lacking in the
Fitzhugh-Nagumo type of systems.

\begin{figure*}[t]
\centering
\hspace{-0.45cm}
\includegraphics[scale=0.209]{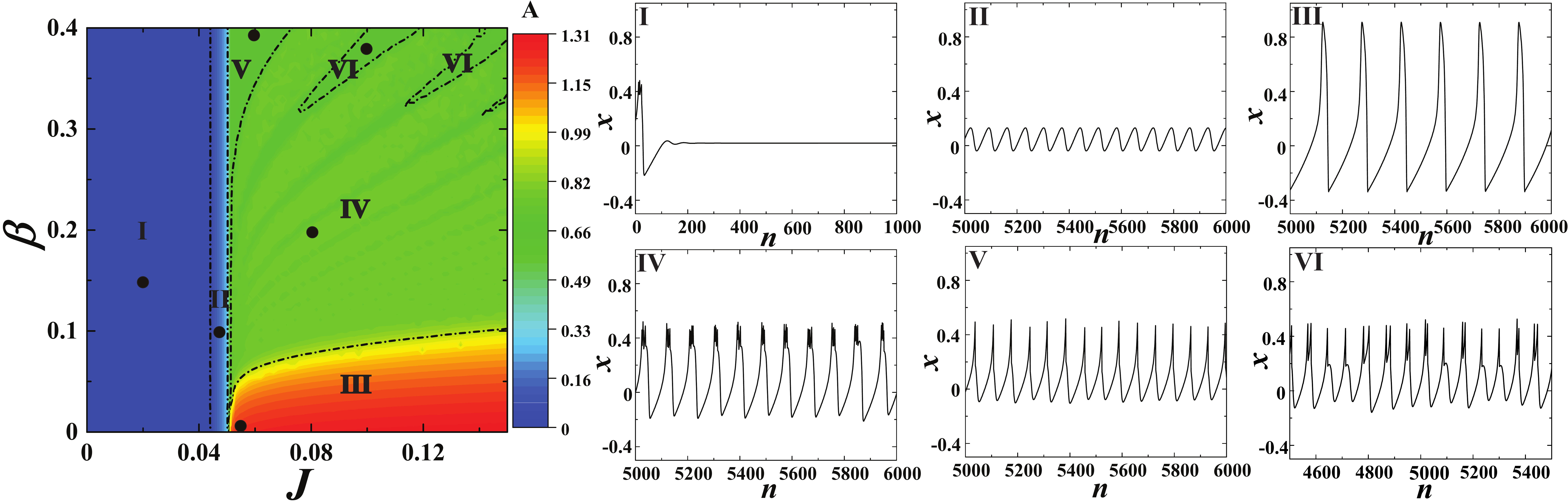}
\caption{(Color online) Dynamical regimes exhibited by model \eqref{eq1}. The heat map refers to variation of the amplitude of oscillations $A$ of the $x$ time series in the $J-\beta$ plane. The waveforms shown in subfigures $I-VI$ illustrate the different forms of neuron's behavior, including excitability $(I)$, subthreshold oscillations $(II)$, regular spiking $(III)$, chaotic bursting ($IV$), chaotic spiking $(V)$, as well as the mixed spike-burst activity $(VI)$. The dots in the heat map indicate the particular $(J,\beta)$ values where the representative waveforms are obtained. \label{Fig1}}
\vspace{-0.4cm}
\end{figure*}

Under variation of $J$ and $\beta$, the map \eqref{eq1} may reproduce a rich repertoire of generic regimes displayed by the real neurons, as demonstrated in Fig. \ref{Fig1}. In particular, the main frame shows amplitudes of the corresponding $x$ time series for the given $(J, \beta)$, while the remaining subfigures illustrate the characteristic waveforms pertaining to excitable regime (region $I$), subthreshold oscillations ($II$), regular ($III$) or chaotic spiking ($IV$), chaotic bursting ($V$), as well as the mixed chaotic spike-burst activity ($VI$). Some of the indicated boundaries, such as those involving domains $IV,V$ and $VI$ should be understood as tentative, since the associated transitions are smooth and therefore difficult to discern.

The detailed phase plane analysis concerning the relevant unstable invariant curves and the mechanisms underlying transitions between the different dynamical regimes can be found in \cite{MN14b}. Here we briefly mention that under increasing $J$, the equilibrium loses stability via the Neimarck-Sacker bifurcation, which gives rise to subthreshold oscillations. Note that the latter may be considered as an excitable state, in a sense that a strong enough perturbation can elicit genuine spike, though the phase point does not relax to the equilibrium, but rather to a closed invariant curve.

\begin{figure}[t]
\centering
\hspace{-0.55cm}
\includegraphics[scale=0.159]{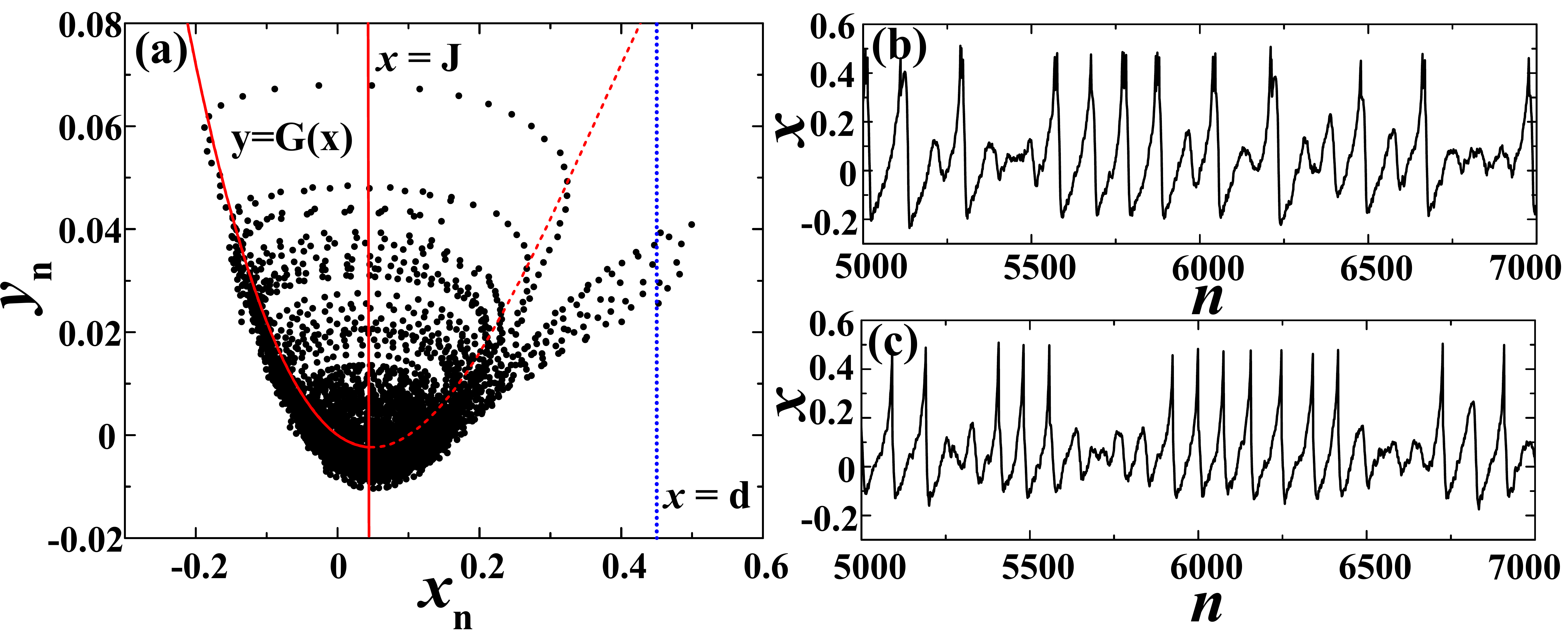}
\caption{(Color online) Impact of noise on a single map neuron in the excitable regime.
(a) indicates the mechanism behind noise-induced spiking. The data are obtained for
$J=0.046,\beta=0.4, \sigma=0.005$. The equilibrium is deterministically stable given that
the line $x=J$ intersects the invariant curve $y=G(x)$ below the curve's minimum.
(b) shows the $x_n$ series corresponding to noise-induced bursting ($J=0.042,\beta=0.2,\sigma=0.008$), whereas (c) demonstrates stochastic spiking superimposed on subthreshold oscillations ($J=0.048,\beta=0.4,\sigma=0.008$).
\label{Fig2}}
\vspace{-0.7cm}
\end{figure}

Adopting model \eqref{eq1} for local dynamics, we focus on a population of $N$ stochastic neurons coupled in the all-to-all fashion via electrical synapses (diffusive couplings). Each neuron receives input from the units within the assembly, and is further influenced by synaptic noise from the embedding environment. The population activity is then described by the following system
\begin{align}\label{eq2}
x_{i,n+1}& = x_{i,n} + G(x_{i,n}) - \beta H(x_{i,n}-d) - y_{i,n} + I_{i,n}^{syn},\\ \nonumber
y_{i,n+1}&= y_{i,n} + \epsilon(x_{i,n} - J), \\
I_{i,n}^{syn}&=I_{i,n}^{coup}+I_{i,n}^{rand}=\frac{c}{N}\sum\limits_{j=1,j\neq i}^N(x_{j,n}-x_{i,n})
+\sigma \xi_{i,n} \nonumber,
\end{align}
where $i$ specifies the particular neuron. The synaptic currents $I_{i,n}^{syn}$ comprise two types of terms. The diffusive couplings $I_{i,n}^{coup}$ are characterized by the strength $c$, which is assumed to be uniform over the network and is set to $c=1$ in the remainder of the paper. The random inputs $I_{i,n}^{rand}$ involve uncorrelated white noise
($E[\xi_{i,n}]=0, E[\xi_{i,n}\xi_{j,n'}]=\delta_{ij}\delta(n-n')$) of intensity $\sigma$.

Confined to a single unit, the stochastic component may influence its dynamics either by perturbing the deterministic oscillatory regimes, or by inducing oscillations in the excitable regime, cf. Fig. \ref{Fig2}(b). The onset of noise-induced spiking or bursting within the parameter domain where the fixed point is deterministically stable (domain $I$ in Fig. \ref{Fig1}) corresponds to a phenomenon of stochastic bifurcation \cite{A99,AB04,GBV11,KSM10,ZSSN05}. The latter are typically described phenomenologically, in a sense that certain time-averaged quantities, such as the asymptotic probability distributions of relevant variables or the associated power spectra, exhibit a qualitative change under variation of noise intensity. For instance, in continuous-time systems, it has been shown that the stochastic Hopf bifurcation from a stochastically stable fixed point to a stochastically stable limit cycle is accompanied by the loss of Gaussian property for the asymptotic distributions of the appropriate variables \cite{TP01}. At variance with standard deterministic bifurcations, where one clearly observes a critical value of the control parameter, the change of system’s behavior in noise-induced transitions is gradual \cite{ZSSN05}. Note that noise can also play an important part in the $(J,\beta)$ region $II$ where the deterministic map shows subthreshold oscillations. Here noise can give rise to a form of dynamics reminiscent of mixed-mode oscillations, cf. Fig. \ref{Fig2}(c).

So far, models similar to \eqref{eq2} have been applied to address a number of problems associated to collective phenomena in networks of coupled neurons, including synchronization of electrically coupled units with spike-burst activity \cite{NM11,CMN12}, pattern formation in complex networks with modular architecture \cite{MKRN13,MN14,MN12}, transient cluster activity in evolving dynamical networks \cite{MN15}, as well as the basin stability of synchronization regimes in small-world networks \cite{MNK15}. Within this paper, the collective motion will be described in terms of the global
variables $X_n=\frac{1}{N}\sum_{i=1}^Nx_{i,n}$ and $Y_n=\frac{1}{N}\sum_{i=1}^Ny_{i,n}$.

\section{Derivation of the mean-field model} \label{sec:MF}

Considering a $MF$ approximation, our main goal lies in deriving a reduced low-dimensional deterministic set of nonlinear difference equations whose dynamics is qualitatively analogous to the collective motion of the original system \eqref{eq2} comprised of $2N$ coupled stochastic maps. In particular, the $MF$ model should be able to generate all the regimes exhibited by the exact system, qualitatively reproducing the bifurcations that the latter undergoes. Also, applying the effective model, one should be capable of inferring with sufficient accuracy the parameter domains which admit the different collective states of the exact system, with the corresponding time series exhibiting similar characteristic quantitative features. Regarding the explicit effects of noise, the $MF$ model is expected to account for the onset or suppression of different types of collective modes associated to macroscopic spiking or bursting activity, which are mediated by synchronization or desynchronization of individual neuron dynamics, respectively. The synchronization processes may be influenced by noise in a variety of ways, including the scenarios where noise acts as a perturbation to mainly deterministic (and chaotic) local oscillations,   or the ones where noise plays a facilitatory role, in a sense that the collective mode emerges via synchronization of noise-induced local dynamics.

Given that we consider a system of discrete-time equations, one cannot adopt the usual method of deriving the $MF$ model via Fokker-Planck formalism \cite{SZNS13}. Nevertheless, an analytically tractable $MF$ model may still be built by focusing on the evolution of cumulants \cite{LGNS04,FTVB14,FTVB13,ZSSN05}, whereby the full density of states is factorized into a series of marginal densities. The advantage of such an approach is that the simplifying approximations aimed at truncating the underlying cumulant series can be introduced in a controlled fashion. Such approximations, stated in a form of closure hypothesis \cite{LGNS04}, are required due to nonlinearity of the original system, which causes the dynamics of cumulants of the given order to be coupled to those of the higher order.

In our case, the derivation of the effective model incorporates an explicit Gaussian closure hypothesis \cite{LGNS04,FTVB14,FTVB13,ZSSN05}, by which all the cumulants above second order are assumed to vanish. The collective dynamics is then described by a set of five variables (the first- and second-order cumulants), including
\begin{itemize}
  \item[(i)] the means, given by $m_{x,n}=\lim\limits_{N\rightarrow\infty}\frac{1}{N}\sum_{i=1}^Nx_{i,n}\equiv\langle x_{i,n}\rangle$, $m_{y,n}=\lim\limits_{N\rightarrow\infty}\frac{1}{N}\sum_{i=1}^Ny_{i,n}\equiv\langle y_{i,n}\rangle$;
  \item[(ii)] the variances, defined as $S_{x,n}=\langle x_{i,n}^2 \rangle-\langle x_{i,n} \rangle^2=
      \langle x_{i,n}^2 \rangle-m_{x,n}^2$ and $S_{y,n}=\langle y_{i,n}^2 \rangle-\langle y_{i,n} \rangle^2=\langle y_{i,n}^2 \rangle-m_{y,n}^2$;
  \item[(iii)] the covariance $U_n=\langle x_{i,n}y_{i,n}\rangle-m_{x,n}m_{y,n}.$
\end{itemize}
The expressions for higher order moments $\langle x_{i,n}^k \rangle$ in terms of the first- and second-order cumulants \cite{G04}, such as \begin{align}
\label{eq3}
\langle x_i^3\rangle&=m_x^3+3m_xS_x \\ \nonumber
\langle x_i^4\rangle&=m_x^4+6m_x^2S_x+3S_x^2 \\ \nonumber
\langle x_i^2y_i\rangle&=m_yS_x+m_ym_x^2+2m_xU \\ \nonumber
\langle x_i^3y_i\rangle&=3S_xU+3S_xm_xm_y+3m_x^2U+m_ym_x^3 \\ \nonumber
\langle x_i^5\rangle&=m_x^5+15m_xS_x^2+10m_x^3S_x \\ \nonumber
\langle x_i^6\rangle&=m_x^6+15S_x^3+15m_x^4S_x+45m_x^2S_x^2,
\end{align}
can be derived using the closure hypothesis.

The Gaussian approximation effectively amounts to an assumption that the relation
\begin{equation}
\lim_{N\rightarrow\infty}\frac{1}{N}\sum_{i=1}^{N}x_{i,n}^k\approx E[x_{i,n}^k], \label{eq4}
\end{equation}
holds, whereby $E$ refers to expectation value obtained by averaging over an ensemble of different stochastic realizations. In other words, one supposes that the local variables are independent and are drawn from a normal distribution $\mathcal{N}(m_x,S_x)$. We do not know \emph{a priori} whether such an assumption is fulfilled or not, but can only judge on its validity by verifying the correctness of the predictions on the population dynamics provided by the $MF$ model. Also note that the effective model concerns the assembly dynamics in the thermodynamic limit $N\rightarrow\infty$. The stochastic terms in this case can be neglected, as one may show them to contribute to finite size effects which scale as $1/N$. This means that the influence of noise in our $MF$ model is felt only via the noise intensity, which assumes the role of an additional bifurcation parameter.

Let us now illustrate the main technical points required for the derivation of the $MF$ model. Our focus will lie with a couple of relevant examples, whereas the remaining details are provided in the Appendix. We begin by considering the dynamics of the global variable $m_x$, which is given by
\begin{equation}
m_{x,n+1}=m_{x,n}-m_{y,n}+\langle G(x_{i,n})\rangle-\beta\langle H(x_{j,n}-d)\rangle \label{eq5}
\end{equation}
It is easy to see that there is no contribution from the coupling term. As far as the third term on the r.h.s. of Eq. \eqref{eq5} is concerned, using Eq. \eqref{eq3}, one arrives at
\begin{equation}
\langle G(x_i)\rangle=\langle -x_i^3+(1+a)x_i^2-ax_i\rangle=G(m_x)+S_x(1+a-3m_x). \label{eq6}
\end{equation} In the last expression, we have dropped the time index for simplicity and
have introduced the shorthand notation $G(m_x)\equiv-m_x^3+(1+a)(m_x^2+S_x)$.

The key problem is how to treat the final term in the r.h.s. of Eq. \eqref{eq5}. Our approach consists
in replacing the assembly average by the expectation value $(\langle H(x_i-d)\rangle\approx E[H(x_i-d)])$, obtained by assuming that the local variables at an arbitrary time moment are normally distributed according to $P(x_i)\sim\mathcal{N}(m_x,S_x)$. The expectation may then be evaluated as
\begin{align}
&E[-\beta\langle H(x_i-d)\rangle]=\int dx_1\int dx_2...\int dx_N \times \\ \nonumber
&(-\frac{\beta}{N}\sum_iH(x_i-d))p(x_1,x_2,...,x_N)=\\ \nonumber
&-\beta\int_{-\infty}^{\infty} dx_1H(x_1-d)p(x_1)
=-\beta\int_{d}^{\infty}\frac{1}{\sqrt{2\pi S_x}}e^{-\frac{(x_1-m_x)^2}{2S_x}}=\\ \nonumber
&-\frac{\beta}{2}(1-Erf\left[\frac{d-m_x}{\sqrt{2S_x}}\right]), \label{eq7}
\end{align}
with the error function $Erf(x)=\frac{2}{\sqrt{\pi}}\int_0^xe^{-t^2}dt$. In
the above calculation, we have explicitly used the assumption on the independence
of distributions of local variables at any given moment of time.

In a similar fashion, one may consider the $S_x$ dynamics, which constitutes the most
demanding part of the derivation. In particular, proceeding from the $S_x$ definition,
we obtain
\begin{align}
S_{x,n+1}&=\langle x_{i,n+1}^2\rangle-\langle x_{i,n+1}\rangle^2=
\langle [(1-c)x_{i,n}+G(x_{i,n})\\ \nonumber
&-\beta H(x_{i,n}-d)-y_{i,n}+\xi_{i,n}+cm_{x,n}]^2\rangle \\ \nonumber
&-(m_{x,n}-m_{y,n}+G(m_{x,n})+S_{x,n}(1+a-3m_{x,n})\\ \nonumber
&-\beta\langle H(x_{i,n}-d)\rangle)^2.\label{eq8}
\end{align}
As an illustration, let us evaluate one of the terms containing an average over
the threshold function:
\begin{align}
&-2\beta E\left[\langle G(x_{i})H(x_{i}-d)\rangle-\right.
\left.\langle G(x_{i})\rangle\langle H(x_{i}-d)\rangle\right]= \\ \nonumber
&-2\beta\left[\int dx_1G(x_1)H(x_1-d)p(x_1)\right. \\ \nonumber
&-\left.\int dx_1H(x_1-d)p(x_1)\left[G(m_x)+S_x(1+a-3m_x)\right]\right]\\ \nonumber
&\approx-2\beta\left[\int dx_1(G(m_x)+G'(m_x)(x_1-m_x)+\frac{1}{2}G''(m_x)\times\right.\\ \nonumber
&(x_1-m_x)^2)H(x_1-d)p(x_1)-\int dx_1H(x_1-d)p(x_1)\times \\ \nonumber
&\left.\left[G(m_x)+S_x(1+a-3m_x)\right]\right]=...=\\ \nonumber
&-2\beta\left[(1+a)(m_x+d)-a-3m_xd\right]\sqrt{\frac{S_x}{2\pi}}\exp\left[-\frac{(d-m_x)^2}{2S_x}\right].
\label{eq9}
\end{align}
Again, the time indexes have been suppressed to simplify the notation.

Leaving the remaining elements of the derivation for the Appendix, we now state the final
equations of the $MF$ model in the thermodynamic limit
\begin{align}
m_{x,n+1}&=m_{x,n}-m_{y,n}+G(m_{x,n})+S_{x,n}(1+a-3m_{x,n}) \\ \nonumber
&-\frac{\beta}{2}(1-Erf\left[\frac{d-m_{x,n}}{\sqrt{2S_{x,n}}}\right]) \\ \nonumber
m_{y,n+1}&=m_{y,n}+\epsilon(m_{x,n}-J) \\ \nonumber
S_{x,n+1}&=(1-c)^2S_{x,n}+S_{y,n}+\sigma^2-2(1-c)U_n\\ \nonumber
&+S_{x,n}(-3m_{x,n}^2+2(1+a)m_{x,n}-a)^2-2(1-c)\times \\ \nonumber
&(3m_{x,n}^2S_{x,n}+3S_{x,n}^2-2(1+a)m_{x,n}S_{x,n}+aS_{x,n})\\ \nonumber
&+2(3S_{x,n}U_n+3m_{x,n}^2U_n-2(1+a)m_{x,n}U_n)\\ \nonumber
&-2\beta\left[(1+a)(m_{x,n}+d)-a-3dm_{x,n}\right]\sqrt{\frac{S_{x,n}}{2\pi}}\times \\ \nonumber
&\exp\left[-\frac{(d-m_{x,n})^2}{2S_{x,n}}\right]
-2\beta(1-c)\sqrt{\frac{S_{x,n}}{2\pi}}\times \\ \nonumber
&\exp\left[-\frac{(d-m_{x,n})^2}{2S_{x,n}}\right]+S_{x,n}^2\left[36m_{x,n}^2-24(1+a)\times\right. \\ \nonumber
&\left.m_{x,n}+2(1+a)^2+6a\right]+15S_{x,n}^3 \\ \nonumber
S_{y,n+1}&=S_{y,n}+\epsilon^2S_{x,n}+2\epsilon U_n \\ \nonumber
U_{n+1}&=U_n-(a+c+\epsilon)U_n+\epsilon(1-c-a)S_{x,n}\\ \nonumber
&-S_{y,n}-(U_n+\epsilon S_{x,n})(3S_{x,n}+3m_{x,n}^2\\ \nonumber
&-2(1+a)m_{x,n})-\beta\epsilon\sqrt{\frac{S_{x,n}}{2\pi}}exp\left[-\frac{(d-m_{x,n})^2}{2S_{x,n}}\right].
\label{eq10}
\end{align}

\section{Analysis of stability and bifurcations} \label{sec:Standbif}

\begin{figure}[t]
\centering
\hspace{-0.7cm}
\includegraphics[scale=0.156]{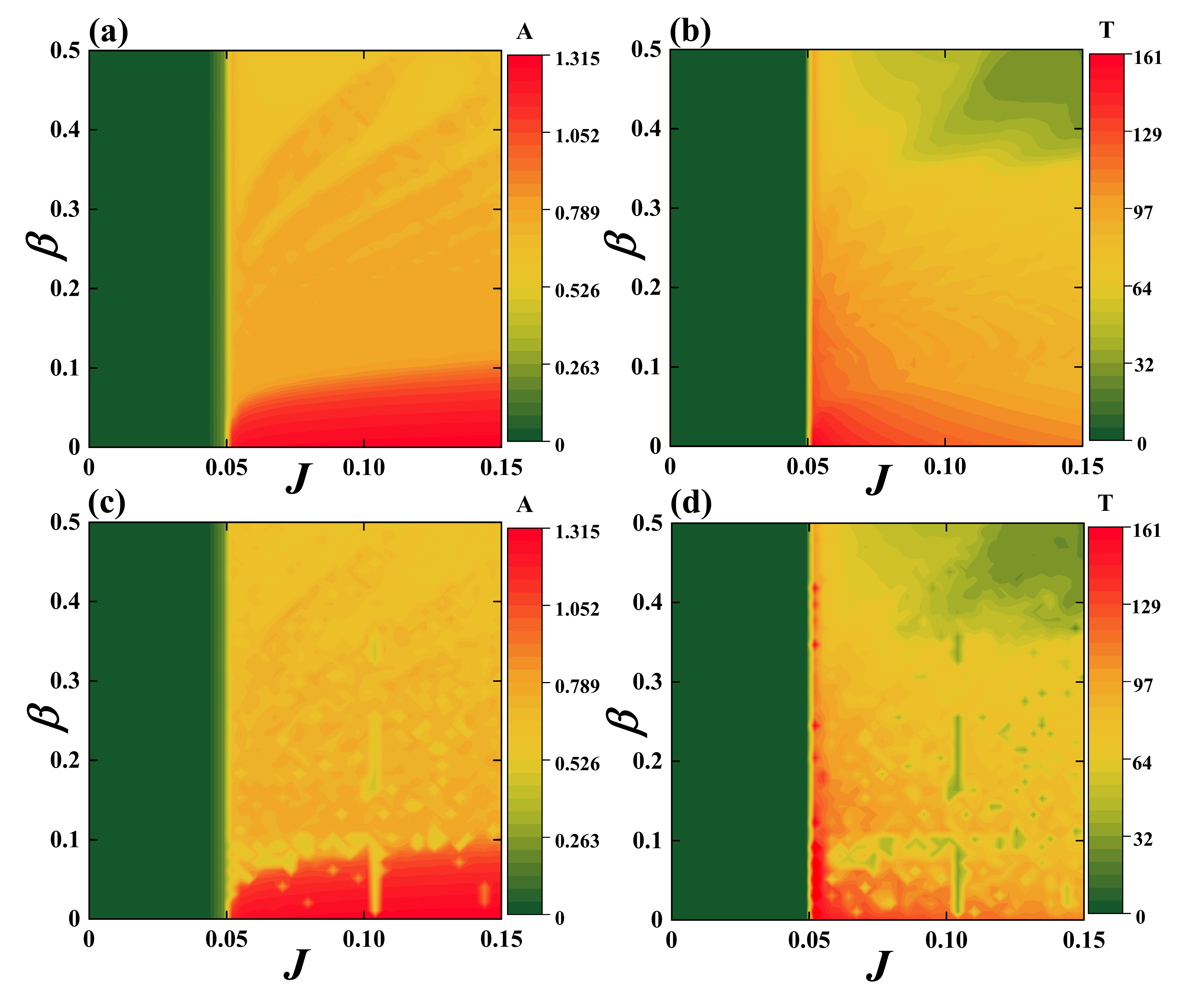}
\vspace{-0.6cm}
\caption{(Color online) Heat maps in (a) and (b) show the dependencies $A(J,\beta)$ and
$T(J,\beta)$ obtained by stochastic averaging for a network of $N=100$ neurons, respectively.
(c) and (d) illustrate the analogous results for the $MF$ model. The noise intensity in
all instances is $\sigma=0.001$. \label{Fig3}}
\vspace{-0.6cm}
\end{figure}

In this section, our goal is to demonstrate the qualitative and quantitative analogies between the dynamics of the exact system and the $MF$ model. To this end, we first examine the succession of macroscopic regimes in the $J-\beta$ parameter plane for $\sigma$ fixed at an intermediate
value $\sigma=0.002$, see Fig. \ref{Fig3}. As in case of a single unit, changing $J$ is relevant for the system’s excitability, viz. the transitions from silent to active regimes, while $\beta$ influences the waveforms of the active states (spiking, bursting, or mixed spike-bursting activity). The assembly is found to exhibit the collective modes which qualitatively correspond to the dynamics of a single unit
illustrated in plates $III-VI$ of Fig. \ref{Fig1}. The heat maps in the left column of Fig. \ref{Fig3} provide a comparison between the oscillation amplitudes $A$ of the global variable $X$ (top row) and the $MF$ variable $m_x$ (bottom row) for the given $(J,\beta)$. The right column indicates how well are matched the average interspike interval (or the average bursting cycle) $T$ of the exact system with the corresponding characteristics of the dynamics of the $MF$ model \eqref{eq10}. In the given instances,  exact system comprises an assembly of $N=100$ neurons, having obtained $A$ by averaging over a sufficiently long time series, whereas $T$ is determined by taking average over an ensemble of $20$ different stochastic realizations. With regard to $T$, we have selected a convenient threshold $\theta=0.2$, which allows a clear detection of individual spikes, and also enables one to unambiguously discern the initiation stage of bursts, as required for calculating the length of the bursting cycle.

\begin{figure}[t]
\centering
\hspace{-0.5cm}
\includegraphics[scale=0.164]{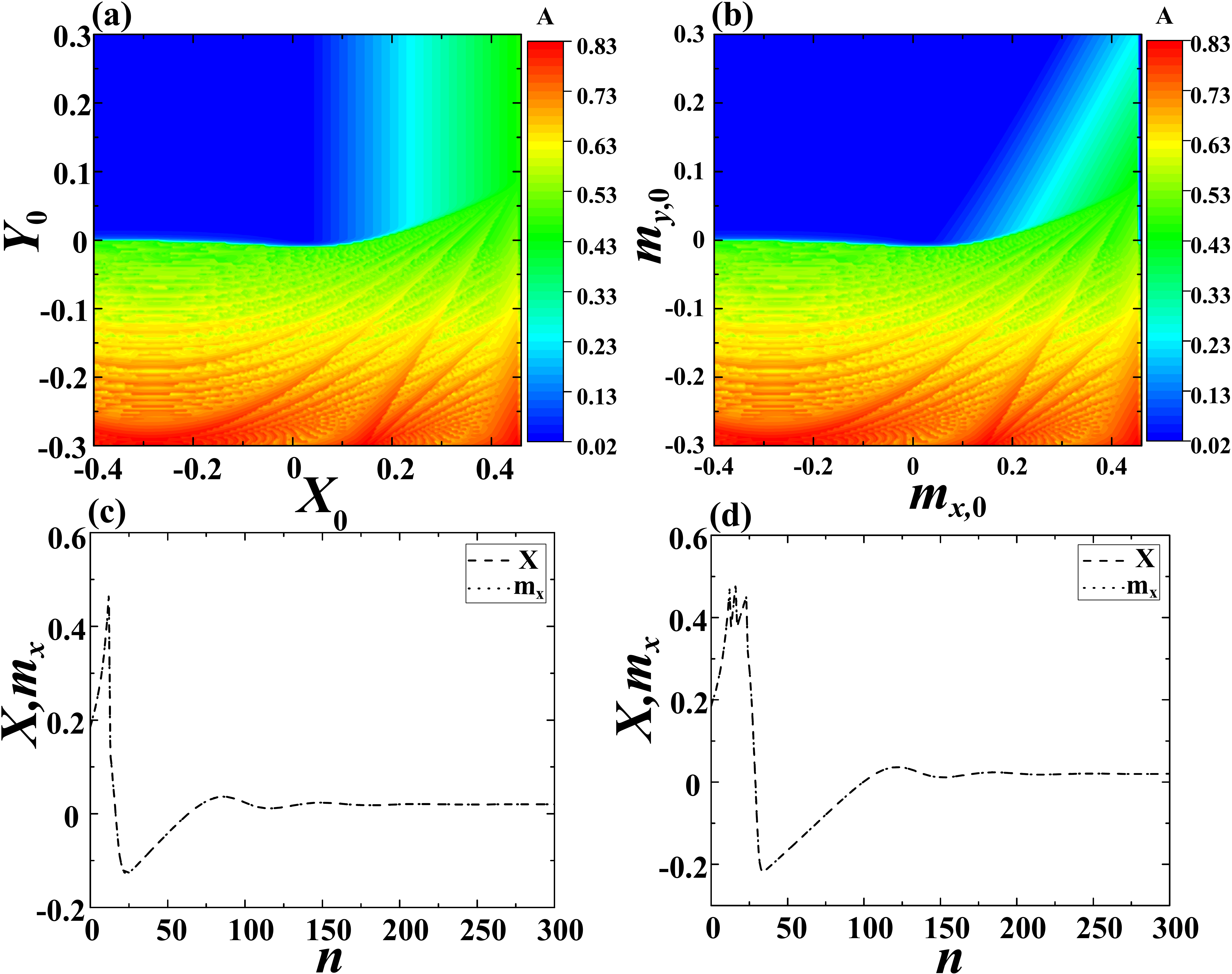}
\vspace{-0.6cm}
\caption{(Color online) Macroscopic excitability feature. In (a) and (b) are shown the maximum
values of $X$ and $m_x$ reached within the time series of the exact and the $MF$ system,
starting from the analogous initial conditions $(X_0,Y_0)$ and $(m_{x,0},m_{y,0})$,
respectively. The parameters are $J=0.02,\beta=0.4$. (c) illustrates the case where
a strong enough perturbation elicits a single-spike response ($J=0.02,\beta=0.4$), whereas
(d) corresponds to a bursting response made up of three spikes ($J=0.02,\beta=0.15$). In
both instances, the time series of the $MF$ model (dotted line) is indistinguishable from
that of the exact system (dashed line). \label{Fig4}}
\vspace{-0.6cm}
\end{figure}

\begin{figure*}[t]
\centering
\includegraphics[scale=0.272]{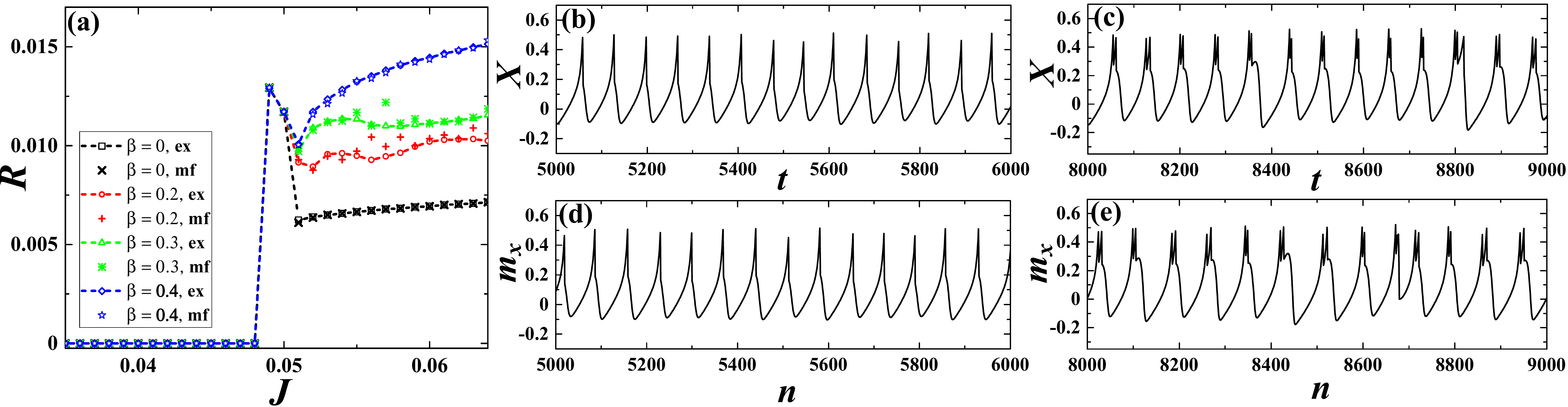}
\vspace{-0.3cm}
\caption{(Color online) (a) shows a family of $R(J)$ curves over $\beta$ for a network of size $N=100$ under fixed $\sigma=0.001$. Superimposed are the results for the $MF$ model, whereby the symbols
$\times,+,\ast,\star$ correspond to cases $\beta=0,0.2,0.3$ and $0.4$, respectively. (b) and (c) illustrate the $X$ series associated to the spiking and the bursting collective modes. The considered network is made up of $N=100$ neurons, with the parameters set to $J=0.06,\beta=0.4,\sigma=0.001$ in (b), and $J=0.08,\beta=0.2,\sigma=0.001$ in (c). In (d) and (e) are provided the $m_x$ series obtained for parameters from (b) and (c). \label{Fig5}}
\vspace{-0.5cm}
\end{figure*}

Let us begin the analysis by focusing on the domain of $J$ values where the exact system exhibits the stochastically stable equilibrium, while the $MF$ model has a stable stationary state. The stochastic stability physically implies that fluctuations around the deterministic fixed point are typically of the order of noise, though some rare spikes may still be evoked. For $J$ sufficiently close to the region admitting the subthreshold oscillations, the population manifests macroscopic excitability. To properly illustrate this feature, we have analyzed the assembly dynamics in the limit $\sigma=0$, cf. Fig. \ref{Fig4}. In particular, figures \ref{Fig4}(a) and \ref{Fig4}(b) show the maximum $X$ and $m_x$ values reached in the corresponding time series obtained for sets of different initial conditions $(X_0,Y_0)$ and $(m_{x,0},m_{y,0})$, respectively. The comparison between the two plots clearly corroborates that the boundary defining the domain of spiking response is appropriately anticipated by the $MF$ model. An important remark is that for the given $J$, the assembly may exhibit different forms of macroscopic excitability, generating a single spike or a burst of spikes, as dependent on the value of $\beta$. This is demonstrated by the time series in figures \ref{Fig4}(c) and \ref{Fig4}(d). The former refers to a one-spike response in case of $\beta=0.4$. For smaller $\beta$, one observes responses comprising two or more closely packed spikes, with Fig. \ref{Fig4}(d) illustrating a three-spike burst encountered for $\beta=0.25$. Note that the time series of the full system and the $MF$ model are exactly matched in the limit $\sigma=0$.

Next we address the noise-influenced transitions from silence to active regimes observed under increasing $J$. To do so, in Figure \ref{Fig5}(a) we have plotted the change of the firing (spiking or bursting) frequency $R$ for an assembly consisting of $N=100$ neurons. The average frequency is determined by considering an ensemble of $20$ different stochastic realizations, having $\sigma$ fixed to the moderate value from Fig. \ref{Fig4}. The results from simulations of the full system \eqref{eq2} are compared against the data obtained for the $MF$ model. In this context, two points should be stressed. First, for moderate $\sigma$, note that the firing frequencies of the $MF$ model lie in close agreement to those of the exact system. As a second point, one finds that such quantitative agreement extends to different forms of collective behavior, viz. it holds for different types of transitions from silent to active regimes. As already indicated, the waveforms pertaining the active states depend on $\beta$, such that the associated transitions are mediated by the distinct synchronization processes. For instance, at $\beta=0$, synchronization involves time series of single units that conform to spiking activity of type $III$ from Figure \ref{Fig1}, which are quite resilient to impact of noise. On the other hand, for $\beta=0.3$ or $\beta=0.4$, the individual units exhibit chaotic bursting or spiking activity, respectively, such that the underlying synchronization process may be more susceptible to stochastic effects. The typical $X$ time series illustrating the different collective modes are compared to the corresponding $m_x$ series in figures \ref{Fig5}(b)-(e). The top (bottom) row concerns the data for the exact system ($MF$ model).

\begin{figure}[b]
\centering
\hspace{-0.4cm}
\includegraphics[scale=0.352]{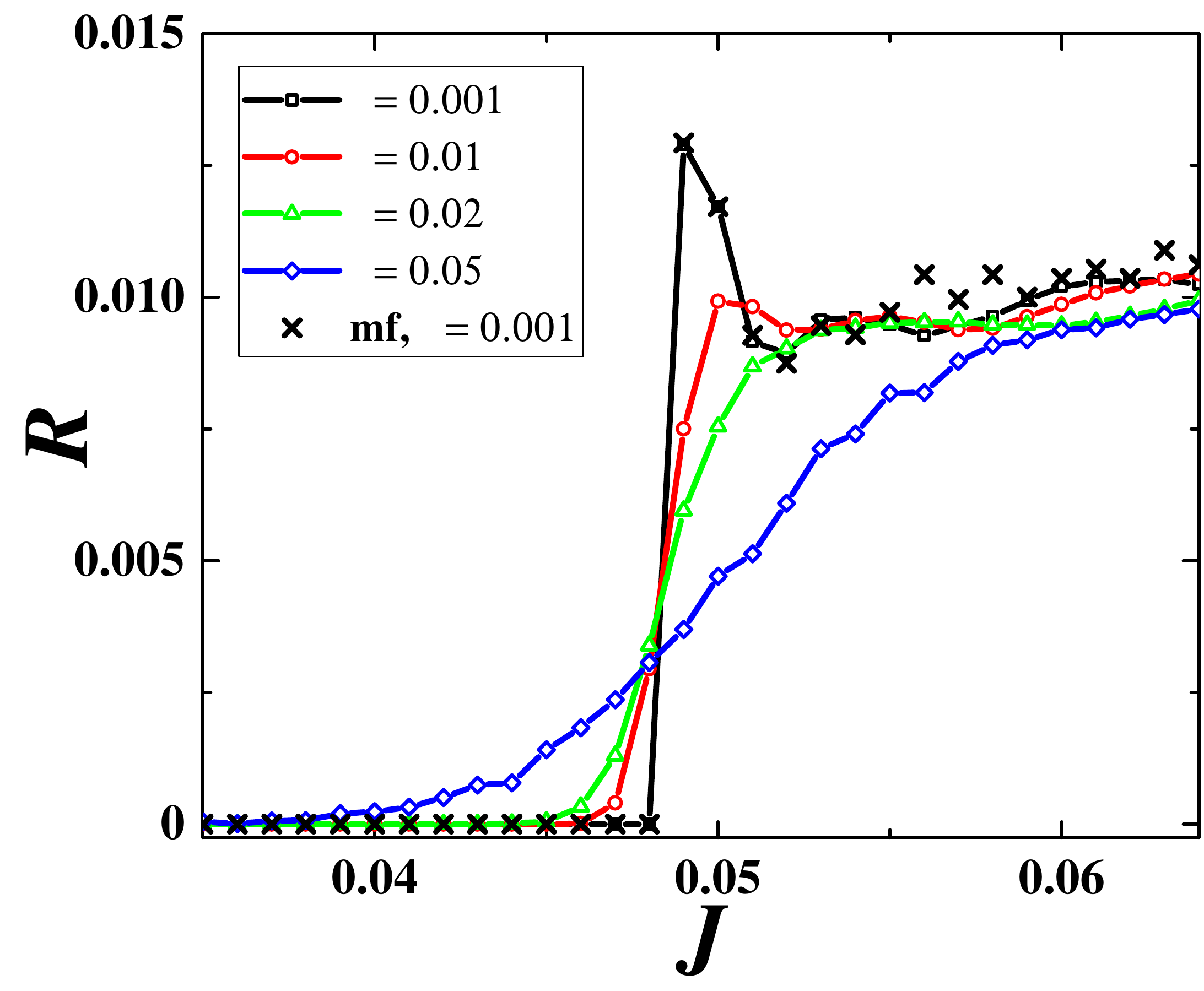}
\vspace{-0.3cm}
\caption{(Color online) Family of $R(J)$ curves over $\sigma$ obtained for a network of $N=100$
neurons under fixed $\beta=0.2$. The different symbols correspond to cases $\sigma=0.001$ (squares),
$\sigma=0.01$ (circles), $\sigma=0.02$ (triangles) and $\sigma=0.05$ (diamonds). The crosses connected
by the dashed line highlight the $R(J)$ curve for the $MF$ model at $\sigma=0.001$.
\label{Fig6}}
\end{figure}

In order to investigate more closely the influence of noise for $J$ interval in vicinity of the transition from silence to active regimes, we examine how the profiles of $R(J)$ curves change under increasing $\sigma$. The results shown in Fig. \ref{Fig6} refer to $\beta=0.2$ and a population comprised of $N=100$ neurons. As expected, the transition appears quite sharp for moderate noise $\sigma=0.001$, but is considerably flattened for larger $\sigma$, e. g. $\sigma=0.05$. The crosses indicate the firing frequencies predicted by the $MF$ model for $\sigma=0.001$.

\begin{figure}[b]
\centering
\hspace{-0.45cm}
\includegraphics[scale=0.143]{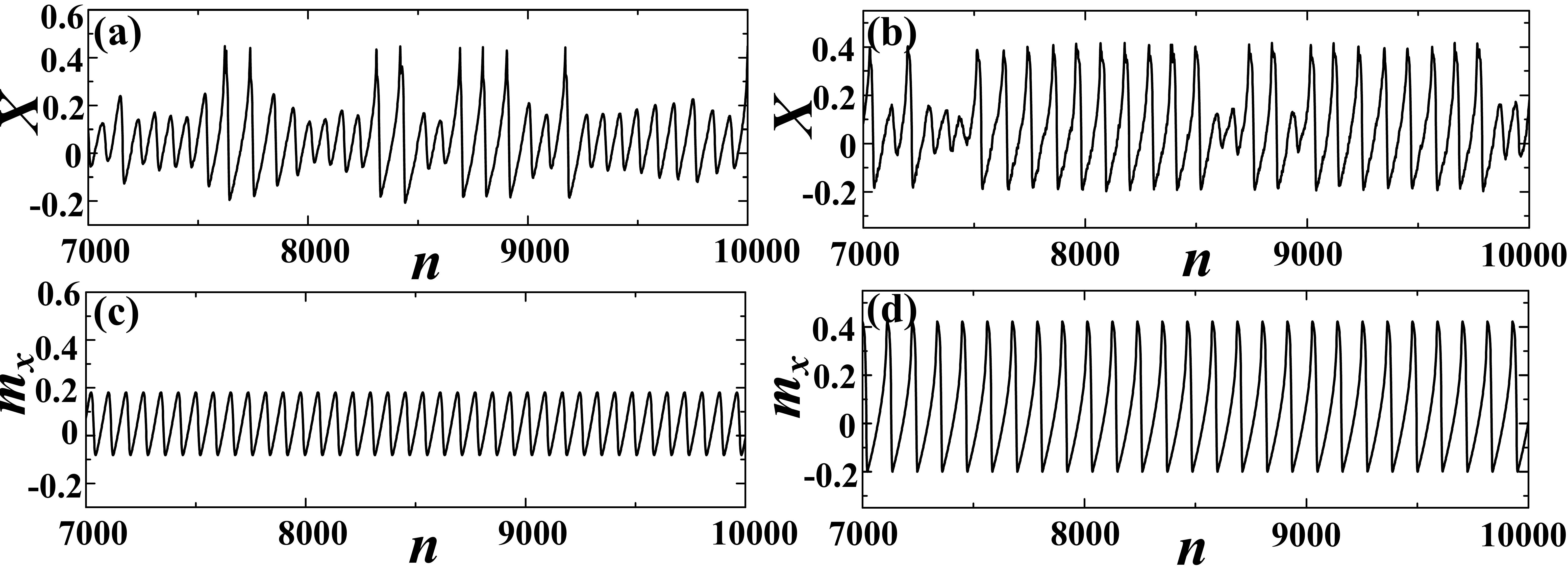}
\caption{Noise-induced phenomena within the $J$ interval in vicinity of the deterministic threshold.
$X$ series in (a) shows the noise-induced spike-bursting activity on top of subthreshold oscillations $(J=0.047,\beta=0.2,\sigma=0.02)$. (b) illustrates the "skipping" phenomenon where the stochastic
effects occasionally suppress the large-amplitude oscillations of the $X$ variable $(J=0.058,\beta=0.2,\sigma=0.01)$. In (c) and (d) are provided the $m_x$ series corresponding to  parameter sets from (a) and (b), respectively. \label{Fig7}}
\end{figure}

For larger $\sigma$, the $MF$ model fails to reproduce the behavior of the exact system in vicinity of threshold $J$, in a sense that it overestimates the maximal $R$ value, as well as the actual critical $J$ characterizing the transition. Viewed from another angle, one may infer that for sufficiently large $\sigma$ and $J$ below the threshold given by the $MF$ model, the latter fails to capture the impact of synchronization processes taking place between the noise-induced oscillations of individual units. This especially refers to $J$ interval where the spikes or bursts (depending on the given $\beta$) are superimposed on the background of subthreshold oscillations. An example of such a discrepancy between the behavior of the exact and the effective system is provided in Fig. \ref{Fig7}, cf. Fig. \ref{Fig7}(a) and Fig. \ref{Fig7}(c). Also, for strong $\sigma$ and $J$ values above the transition, the firing frequencies anticipated by the effective model are typically higher than those of the exact system (not shown). Within this region, the stochastic effects suppress synchronization between the chaotic oscillations of single neurons, thereby reducing the corresponding $R$ value. This is not accounted for with sufficient accuracy by the $MF$ system. Note that such suppression of synchronization is reflected in the corresponding $X$ series by the spike (burst) "skipping" mechanism, where the large-amplitude oscillations are occasionally replaced with subthreshold oscillations. For the associated $J$ and $\sigma$ values, such a phenomenon is absent in the dynamics of the effective model, cf. Fig. \ref{Fig7}(b) and Fig. \ref{Fig7}(d). In both of the scenarios illustrated in Fig. \ref{Fig7}, the reason for having the $MF$ model fail lies in that the Gaussian approximation behind it breaks down due to large stochastic fluctuations.

\begin{figure}[t]
\centering
\hspace{-0.6cm}
\includegraphics[scale=0.173]{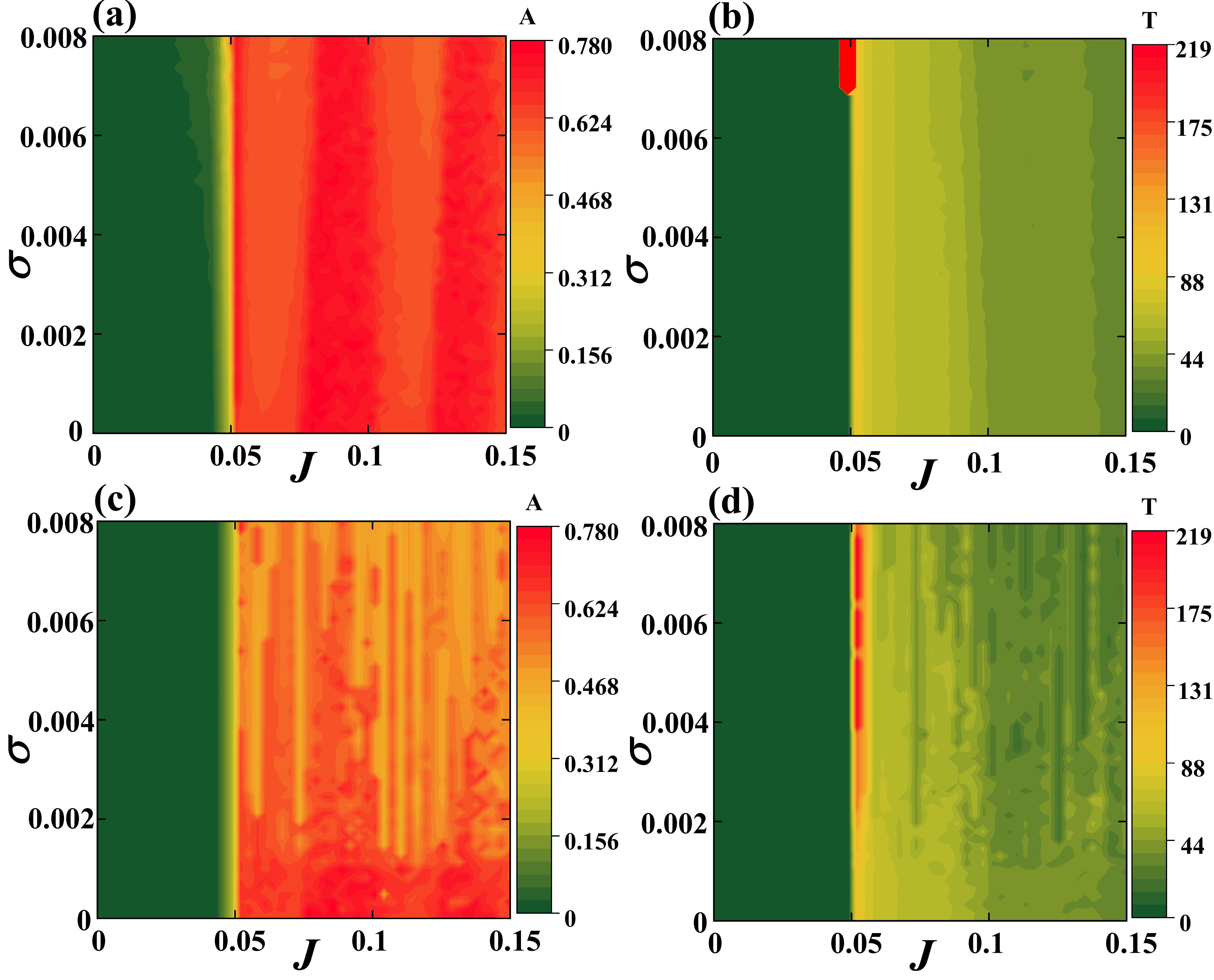}
\vspace{-0.5cm}
\caption{(Color online) (a) and (b) respectively refer to $A(J,\sigma)$ and $T(J,\sigma)$
dependencies for the network of $N=100$ neurons under fixed $\beta=0.4$. The results in (a)
are obtained by averaging over a sufficiently long time series, whereas data in (b) derive
from averaging over an ensemble of $20$ different stochastic realizations. In (c) and (d)
are provided the $A(J,\sigma)$ and $T(J,\sigma)$ dependencies determined by numerical simulations
of the $MF$ model.
\label{Fig8}}
\end{figure}

The fashion in which the validity of the effective model’s predictions deteriorates with increasing $\sigma$ is made more explicit in Fig. \ref{Fig8}, which shows the $A(J,\sigma)$ and $T(J,\sigma)$ dependencies for the exact and the approximate system at fixed $\beta=0.4$. The considered size of the network is $N=100$. Comparison between the respective $A$ (left column) and $T$ plots (right column) suggests that the range of $\sigma$ values where the $MF$ approximation applies is contingent on $J$. For instance, in the $J$ region below the deterministic threshold, one may estimate this range by noting that the effective bifurcation diagram in Fig. \ref{Fig8}(a) indicates that noise-induced macroscopic oscillations emerge for $\sigma\approx0.003$. Since this point is not adequately represented by the effective model, cf. Fig. \ref{Fig8}(c), one may state that the Gaussian approximation breaks down around $\sigma\approx0.003$ within the given $J$ region. Nevertheless, for $J$ above the deterministic threshold, the validity of the $MF$ model appears to depend rather strongly on particular $J$, with the $\sigma$ values where the Gaussian approximation effectively fails spanning the range $\sigma\in(0.002,0.006)$.

\begin{figure}[b]
\centering
\hspace{-0.4cm}
\includegraphics[scale=0.365]{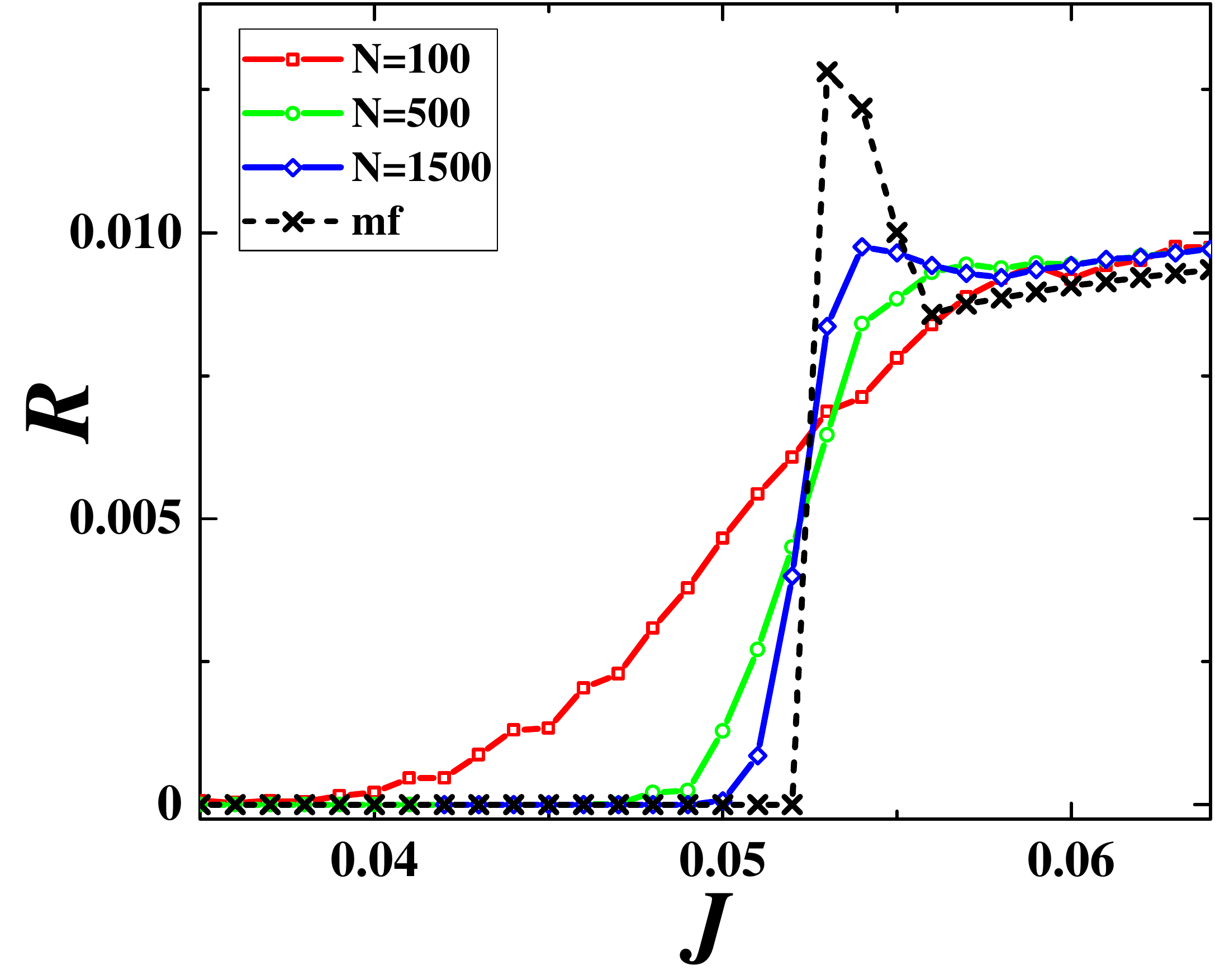}
\vspace{-0.4cm}
\caption{(Color online) $R(J)$ dependencies for increasing $N$ under fixed $(\beta,\sigma)=(0.2,0.05)$.
The squares, circles and diamonds correspond to cases $N=100$, $N=500$ and $N=1500$, respectively. The
results predicted by the $MF$ model are indicated by crosses connected via dashed line.
\label{Fig9}}
\end{figure}

So far, we have investigated the impact of noise by comparing the results for the network of size $N=100$ to those obtained for the effective system. Nevertheless, within Section \ref{sec:MF}, it has already been emphasized that the $MF$ model, deterministic in character, refers to the system’s behavior in the thermodynamic limit $N\rightarrow\infty$, whereas the explicitly stochastic terms could only be incorporated as finite-size effects. This makes it relevant to examine how the behavior of the exact system within the $J$ domain around deterministic threshold changes for large and fixed $\sigma$ under increasing $N$. To this end, we have plotted in Fig. \ref{Fig9} the $R(J)$ curves calculated for $N=100$ (squares), $N=500$ (circles) and $N=1500$ (diamonds) at fixed $\beta=0.2,\sigma=0.05$. The curve for $N=100$ evinces that the given $\sigma$ value is quite large in a sense of being sufficient to induce collective oscillations within the excitable regime. Apart from the dependencies for the full system, we also show the $R(J)$ curve associated to the $MF$ model (dashed line with crosses). An interesting point regarding the latter is that the $J$ threshold for the emergence of the collective mode is shifted toward a larger value compared to the case $\sigma\approx 0.01$. While the given transition itself appears quite sharp, the curves corresponding to the exact system approach it with increasing $N$, both in terms of the $J$ threshold and the $R$ values above the transition. This corroborates that the $(J,\sigma)$ domain where the Gaussian approximation behind the $MF$ model fails expectedly reduces with the increasing system size.

\section{Response to external stimuli} \label{sec:Responses}

The aim of this section is to investigate the extent to which the $MF$ model can be used to predict the stimulus-response relationship of an assembly exhibiting different macroscopic regimes, including the excitable state, as well as the spiking and bursting collective modes. Let us first focus on the two latter instances and examine the sensitivity of a population to an external pulse perturbation within the framework of phase resetting theory \cite{SPB12,T07,I07,C06}. In order to compare the behavior of the exact system and the effective model, we determine the corresponding phase resetting curves ($PRC$s), which describe the phase shift $\Delta\phi$, induced by the perturbation, in terms of the phase $\phi_p$ when the perturbation is applied. The considered stimulus has a form of a short pulse current $I_p=a_pH(n-n_i)H(n-n_f)$, whose magnitude $a_p$ and width $\Delta=n_i-n_f$ are small compared to the amplitude and duration of the spiking (or bursting) cycle $T_0$, respectively.  In case of the exact system, the same pulse current is delivered to each neuron $i$, adding the term $I_p$ to $x_i$ dynamics, whereas in the effective model, stimulation is administered via the $m_x$ variable. The phase $\varphi_p$ is defined in reference to $T_0$ by $\varphi_p=n_p/T_0$. The associated phase difference following the reset is calculated as $\Delta\varphi=1-T_1/T_0$, where $T_1$ denotes the duration of the perturbed spiking or bursting cycle.

\begin{figure}[b]
\centering
\includegraphics[scale=0.145]{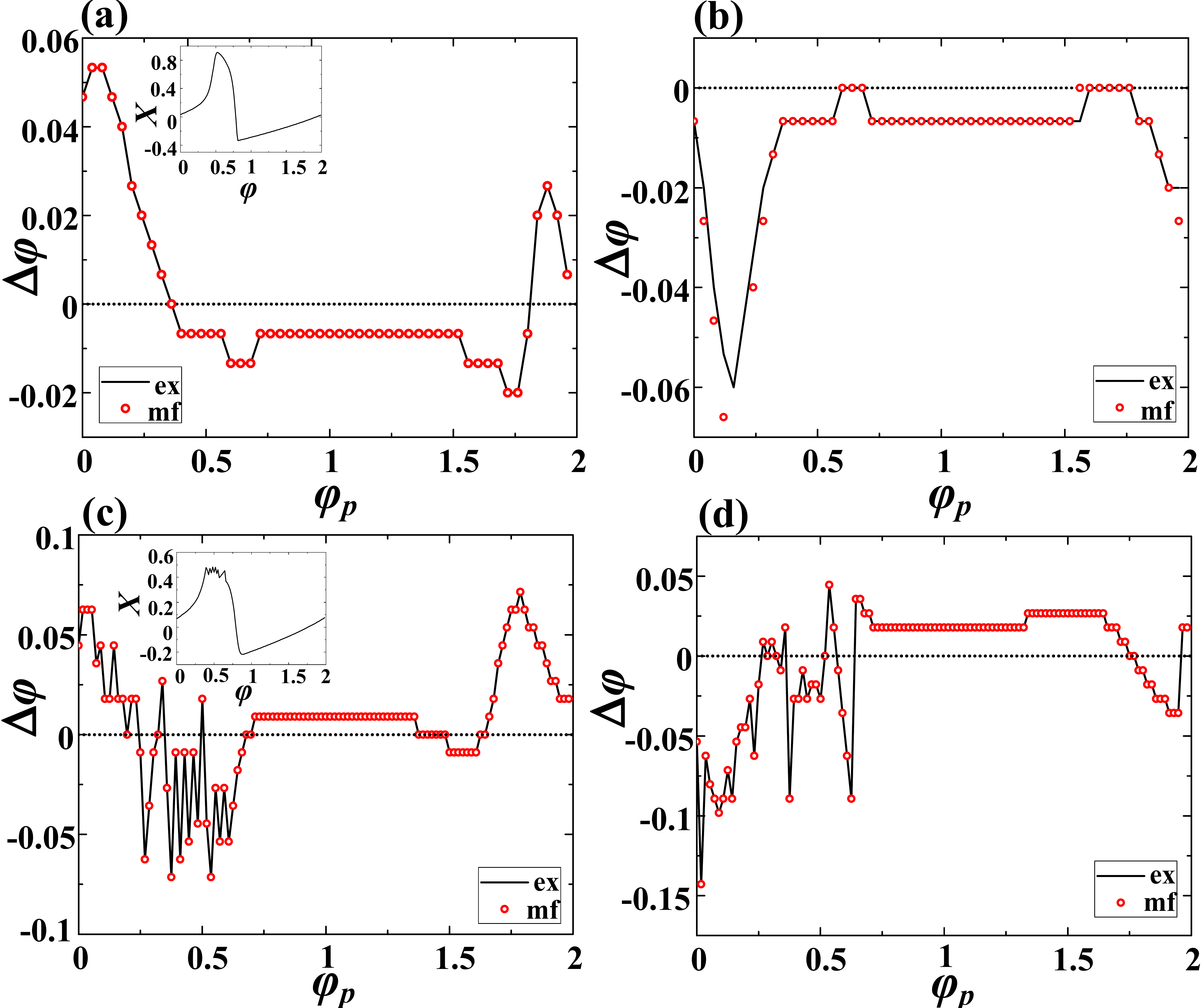}
\caption{(Color online) Assembly phase resetting. (a) and (b) show the $PRC$s for a population in spiking regime ($J=0.055,\beta=0$) under excitatory ($a=0.008$) and inhibitory stimulation ($a=-0.008$), respectively. Results for the exact system ($N=500$) are indicated by the solid line, whereas the data for the $MF$ model are denoted by circles. The bottom row illustrates the $PRC$s for an assembly exhibiting macroscopic bursting ($J=0.06,\beta=0.1$), whereby (c) describes the effect of an excitatory ($a=0.01$), and (d) of an inhibitory pulse perturbation ($a=-0.01$). The insets in (a) and (c) demonstrate how the phases are assigned to the points within the spiking and bursting cycles, respectively. Phase is expressed in units of $\pi$.
\label{Fig10}}
\end{figure}

\begin{figure*}[t]
\centering
\includegraphics[scale=0.195]{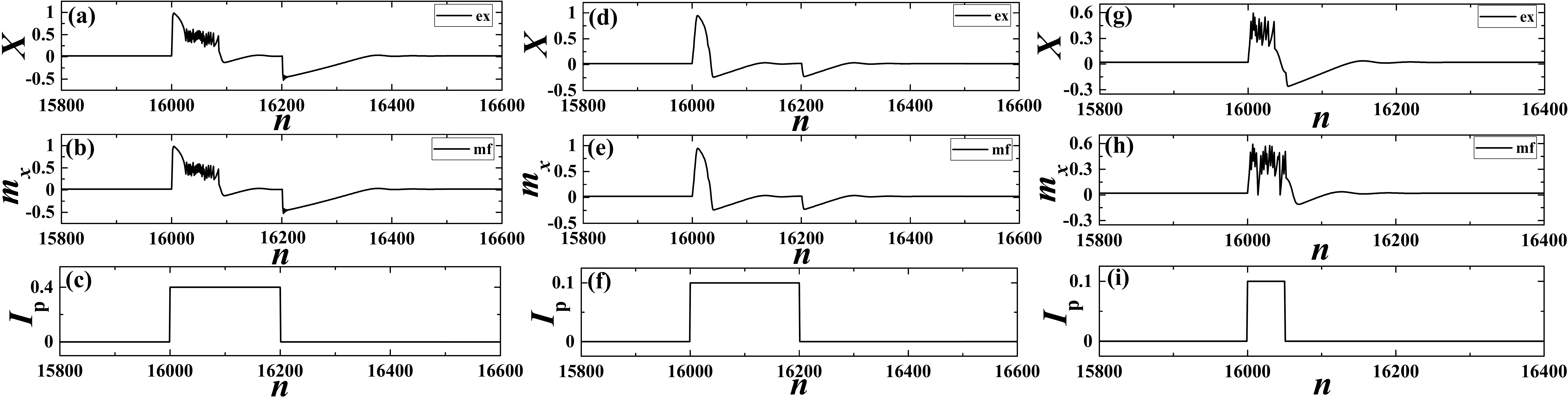}
\vspace{-0.3cm}
\caption{Stimulus-response relationship in the excitable regime ($J=0.02$). The top (middle) row refers to the response of the full system ($MF$ model), whereas the bottom row shows the profile of the external stimulation. In the panels (a)-(c), the system parameters are $\beta=0.4,\sigma=0$, while the perturbation is characterized by $a_p=0.4,\Delta=200$. Panels (d)-(f) concern the response of an assembly ($\beta=0.1,\sigma=0.001$) subjected to a rectangular pulse $a_p=0.4,\Delta=200$. Panels (g)-(i) illustrate the response of a population ($\beta=0.4,\sigma=0.001$) influenced by the external stimulation $a_p=0.1,\Delta=50$. The considered network is of size $N=500$.
\label{Fig11}}
\vspace{-0.5cm}
\end{figure*}

The $PRC$s characterizing the assembly response in the spiking regime are provided in Fig. \ref{Fig10}(a) and Fig. \ref{Fig10}(b), whereby the former is obtained under the action of an excitatory ($a_p>0$), and the latter under the influence of an inhibitory stimulation ($a_p<0$). We stress that in both instances, the results derived from the effective model, denoted by circles, show excellent agreement with the data for the exact system (solid lines). In qualitative terms, one observes that an excitatory stimulation may advance the phase of the spiking cycle if it arrives sufficiently close to the spike, but still before the sharp rising stage. Nevertheless, an excitatory perturbation which acts during the spike or within the effective refractory period has a suppression effect, reflected in delaying of the next spike. At variance with the excitatory stimulation, the inhibitory pulse postpones the next firing time if it is introduced within the interval close to the rising stage of spike.

The $PRC$s determined for an assembly exhibiting collective bursting show qualitatively analogous effects to those described so far, see Fig. \ref{Fig10}(c) and Fig. \ref{Fig10}(d). This especially refers to impact of perturbation delivered sufficiently close to a moment of burst initiation. An apparent difference compared to Fig. \ref{Fig10}(a) and Fig. \ref{Fig10}(b) emerges during the bursting stage itself, where the associated $PRC$s expectedly exhibit strong fluctuations. Apart from that, one finds an interesting effect that both the excitatory and the inhibitory stimulation have a facilitatory role,
i. e. cause phase advancement during the relaxation stage of the bursting cycle.

For a population in the excitable state, we consider scenarios where the system is influenced by a rectangular pulse perturbation of finite magnitude and duration, in a sense that the latter are comparable to corresponding features of typical spiking or bursting cycles. Note that the selected $J$ value $J=0.02$ lies sufficiently away from the interval admitting the subthreshold oscillations. Again, our objective is to determine whether the $MF$ model correctly anticipates the response of the exact system, now in presence of small to moderate noise. Some of the illustrative examples concerning the   stimulus-response relationship under the finite perturbation are provided in Fig. \ref{Fig11}. The top and the middle row refer to $X$ and corresponding $m_x$ time series, respectively, while the bottom row shows the profile of the applied stimulus. We find that in the absence of noise or for sufficiently small $\sigma$, the effective model reproduces the evoked behavior of the full system quite accurately. This also refers to some highly complex forms of responses, as corroborated in Fig. \ref{Fig11}(a)-(c), which concern relatively large $a_p$ and $\Delta$. Under increasing $\sigma$, the ability of the $MF$ model to predict the dynamics of the exact system gradually reduces, but in a fashion that involves a nontrivial dependence on $\beta$. In particular, for smaller $\beta\approx0.1$, which would facilitate macroscopic spiking mode for supercritical $J$, it turns out that the dynamics of the $MF$ model lies in close agreement to the one of the exact system even for moderate noise $\sigma=0.001$, cf. Fig. \ref{Fig11}(d)-(f). However, for higher $\beta$, such an analogy between the responses of the exact and the $MF$ system is lost, see Fig. \ref{Fig11}(g)-(i). Naturally, the validity of the predictions given by the $MF$ model deteriorates if the stimulation amplitude $a_p$ and the duration $\Delta$ are large, especially in presence of non-negligible noise.

\section{Summary and discussion} \label{sec:Summary}

We have developed a $MF$ approach in order to systematically analyze the emergent dynamics and the input-output relationship of a population of stochastic map neurons. The reduced low-dimensional model has been derived within the framework of Gaussian approximation, formally introduced in a form of a closure hypothesis. In physical terms, such an approximation suggests that the local variables at an arbitrary moment of time are independent and conform to a normal distribution centered about the assembly mean and characterized by the associated assembly variance. Validity of such an approximation cannot be established \emph{a priori}, but has been systematically verified by numerically corroborating that the $MF$ model reproduces the behavior of the exact system with sufficient accuracy.

In particular, we have first demonstrated that the effective model can qualitatively capture all the bifurcations of the exact system leading to the onset of different generic regimes of collective behavior. As far as the quantitative agreement is concerned, we have established substantial matching between the parameter domains admitting the respective dynamical regimes for the exact and the approximate system. Moreover, the typical features of the associated regimes, such as the average interspike interval or the average bursting cycle, exhibit analogous changes with parameter variation, and in many parameter domains display numerically similar values.

An important issue has been to explicitly examine how the effects of noise are reflected in the behavior of the $MF$ model. For the noise-perturbed activity, where the sufficiently small noise weakly influences the deterministic attractors of the system, the obtained results indicate that the Gaussian approximation holds. Nevertheless, the physical picture changes in case of noise-induced collective behavior. In particular, for different scenarios of stochastic bifurcations, typically corresponding to transitions from subthreshold oscillations, which involve generalized excitability feature, to spiking or bursting regimes, the exact system undergoes a gradual (smooth) change of collective dynamics, whereas the $MF$ model exhibits a standard deterministic bifurcation with a sharp bifurcation threshold. In such instances, the collective variables of exact system manifest large fluctuations, which explicitly violate the Gaussian approximation behind the effective model. Note that the loss of Gaussianity property for asymptotic distribution of relevant variables, which accompanies the described stochastic bifurcations, does not imply \emph{per se} that our Gaussian approximation fails in the supercritical state. This point is evinced by the fact that the dynamics of the effective model shows qualitatively and quantitatively similar features to those of the exact system if the considered parameters lie sufficiently above the stochastic bifurcation. In fact, the Gaussian approximation applied in the derivation of the $MF$ model breaks down only in vicinity of such transitions, where the finite-size effects neglected in Eq. \eqref{eq10} become most prominent. We have numerically verified the prevalence of finite-size effects in these parameter domains, showing that the change of the appropriate order parameter, such as the spiking frequency, becomes sharper as the size of the neural assembly is increased. Nevertheless, the validity of Gaussian approximation is regained once the system is sufficiently above the bifurcation.

Apart from considering asymptotic dynamics, we have verified that the $MF$ model is capable of capturing the stimulus-response features of the exact system. For short pulse-like perturbations, it has been found that the approximate system reproduces the $PRC$s of the exact system for both the spiking and bursting regimes of collective activity with high accuracy. Substantial analogies have also been observed in case of macroscopic excitable regime for scenarios where the assembly is stimulated by rectangular pulse perturbations of finite amplitude and duration.

Having developed a viable $MF$ approach, the present research has set the stage for a more systematic exploration of collective dynamics of assemblies of map neurons by analytical means. We believe that the introduced techniques can be successfully applied for treating the emergent behavior of populations in case of chemically and delay-coupled neurons \cite{MN14}. Moreover, the method may likely be used to explore the effects of parameter inhomogeneity, as well as to study the impact of complex network topologies \cite{MN14,MNK15}. Our ultimate goal will be to extend the $MF$ approach to account for collective behavior of interacting populations of map neurons \cite{MKRN13,MN14}.

\begin{acknowledgments}
This work is supported by the Ministry of Education, Science and Technological
Development of Republic of Serbia under project No. $171017$, and by the Russian
Foundation for Basic Research under project No. 15-02-04245.
\end{acknowledgments}

\section{Appendix}

In the following, we provide the remaining details concerning the calculation of the $S_x$ dynamics, which is the most complex part of the derivation of the effective model. Following some algebra, Eq. \eqref{eq8} can be transformed to
\begin{align}
S_{x,n+1}&=(1-c)^2S_{x,n}+S_{y,n}+\sigma^2-2(1-c)U_n \\ \nonumber
&+\underbrace{(\langle G(x_{i,n})^2\rangle-\langle G(x_{i,n})\rangle^2)}_{Var(G(x_{i,n}))}+2(1-c)
(\langle x_{i,n}G(x_{i,n})\rangle \\ \nonumber
&-m_{x,n}\langle G(x_{i,n})\rangle)-2(\langle y_{i,n}G(x_{i,n})\rangle-m_{y,n}\langle G(x_{i,n})\rangle)\\ \nonumber
&-2\beta(1-c)
\left[\langle x_{i,n}H(x_{i,n}-d)\rangle-m_{x,n}\langle H(x_{i,n}-d)\rangle\right] \\ \nonumber
&-2\beta(\langle G(x_{i,n})H(x_{i,n}-d)\rangle-\langle G(x_{i,n})\rangle\langle H(x_{i,n}-d)\rangle)\\ \nonumber
&+\beta^2\underbrace{(\langle H(x_{i,n}-d)^2\rangle-\langle H(x_{i,n}-d)\rangle^2)}_{Var(H(x_{i,n}-d)))}.
\label{eqa1}
\end{align}

The partial results required for completing the calculation are given by
\begin{align}
\langle x_iG(x_i)\rangle-m_x\langle G(x_i)\rangle&=G'(m_x)S_x-3S_x^2 \\ \nonumber
\langle y_iG(x_i)\rangle-m_y\langle G(x_i)\rangle&=-3S_xU_{xy}-3m_x^2U_{xy} \\ \nonumber
&+2(1+a)m_xU_{xy},
\label{eqa2}
\end{align}
where $G'(m_x)\equiv-3m_x^2+2(1+a)m_x-a$. Note that the time indexes have been omitted for simplicity.
After some tedious work, it may also be shown that the expression for variance $Var(G(x_i))$ reads
\begin{align}
Var(G(x_i))&=G'^2(m_x)S_x+S_x^2\left[36m_x^2-24(1+a)m_x\right. \\ \nonumber
&+\left.2(1+a)^2+6a\right]+15S_x^3. \label{eqa3}
\end{align}

Let us now explicitly calculate the terms containing the threshold function. First we have
\begin{align}
&-2\beta(1-c)
\left[\langle x_iH(x_i-d)\rangle-\langle x_i\rangle\langle H(x_i-d)\rangle\right]=-2\beta\times\\ \nonumber
&(1-c)\left[\int dx_1dx_2...dx_N\frac{1}{N}\sum_i x_iH(x_i-d)p(x_1,...,x_N)\right. \\ \nonumber
&\left.-m_x\int dx_1dx_2...dx_N \frac{1}{N}\sum_i H(x_i-d)p(x_1,...,x_N)\right]=\\ \nonumber
&...=-2\beta(1-c)\left[\int dx_1(x_1-m_x)H(x_1-d)p(x_1)\right]=\\ \nonumber
&-2\beta(1-c)\sqrt{\frac{S_x}{2\pi}}\exp\left[-\frac{(d-m_x)^2}{2S_x}\right]. \label{eqa4}
\end{align}
Note that the second term containing the threshold function has been evaluated in the main text,
cf. Eq. \eqref{eq9}.

Finally, let us address the term $ \beta^2Var(H(x_i-d))$, which can be estimated by
considering the associated expectation $\beta^2Var(H(x_i-d))\approx\beta^2\left[\langle H(x_i-d)^2\rangle-\langle H(x_i-d)\rangle^2\right]$. Applying the technique  introduced in Sec. \ref{sec:MF}, we obtain
\begin{align}
&E[\beta^2H(x_i-d)^2]=\beta^2\int dx_1\int dx_2...\int dx_N\times \\ \nonumber
&(\frac{1}{N^2}\sum_i\sum_jH(x_i-d)H(x_j-d))p(x_1,x_2,...,x_N) \\ \nonumber
&=\underbrace{\frac{\beta^2}{N^2}N\int dx_1H(x_1-d)p(x_1)}_{N\,\,cases\,\,where\,\,i=j} \\ \nonumber
&+\underbrace{\frac{\beta^2}{N^2}N(N-1)\int dx_1\int dx_2H(x_1-d)H(x_2-d)p(x_1)p(x_2)}_{{N(N-1)\,\,cases\,\,where\,\,i\neq j}} \\ \nonumber
&=\frac{\beta^2}{2N}[1-Erf[\frac{d-m_x}{\sqrt{2S_x}}]]+
\frac{\beta^2}{4N^2}N(N-1)[1-Erf[\frac{d-m_x}{\sqrt{2S_x}}]]^2.\label{eqa5}
\end{align}
Given that $\beta^2\langle H(x_i-d)\rangle^2=\frac{\beta^2}{4}\left[1-Erf[\frac{d-m_x}{\sqrt{2S_x}}]\right]^2$,
one arrives at
\begin{equation}
\beta^2Var(H(x_i-d))=\frac{\beta^2}{4N}(1-Erf[\frac{d-m_x}{\sqrt{2S_x}}])
([1+Erf[\frac{d-m_x}{\sqrt{2S_x}}]). \label{eqa6}
\end{equation}
This shows that the variance of the threshold function ultimately contributes to a finite-size effect which can be neglected in the thermodynamic limit.


\begin{thebibliography}{99}

\bibitem{ZZHK06}{C. Zhou, L. Zemanova, G. Zamora, C. Hilgetag, and J. Kurths,
Phys. Rev. Lett. \textbf{97}, 238103 (2006).}

\bibitem{BS09}{E. Bullmore, and O. Sporns, Nat. Rev. Neurosci. \textbf{10}, 186 (2009).}

\bibitem{MLB10}{D. Meunier, R. Lambiotte, and E. Bullmore, Front. Neurosci. \textbf{4},
200 (2010).}

\bibitem{SCKH04}{O. Sporns, D. Chialvo, M. Kaiser, and C. C. Hilgetag,
Trends Cogn. Sci. \textbf{8}, 418 (2004).}

\bibitem{MPS10}{\emph{Dynamic Coordination in the Brain}, edited by C. von der Malsburg,
W. A. Phillips, and W. Singer, (MIT Press, Cambridge, 2010).}

\bibitem{ALP06}{B. B. Averbeck, P. E. Latham, and A. Pouget,
Nat. Rev. Neurosci. \textbf{7}, 358 (2006).}

\bibitem{VLRM01}{F. Varela, J. P. Lachaux, E. Rodriguez, and J. Martinerie,
Nat. Rev. Neurosci. \textbf{2}, 229 (2001).}

\bibitem{B09}{G. Buzs\'{a}ki, \emph{Rhythms of the Brain}, (Oxford University Press, Oxford, 2009).}

\bibitem{VW09}{J. L. P. Velazquez, and R. Wennberg, \emph{Coordinated Activity in the Brain:
Measurements and Relevance to Brain Function and Behavior}, (Springer, New York, 2009).}

\bibitem{CS06}{R. T. Canolty et al., Science \textbf{313}, 1626 (2006).}

\bibitem{LJ13}{J. E. Lisman, and O. Jensen, Neuron \textbf{77}, 1002 (2013).}

\bibitem{BRZK09}{Y. Baibolatov, M. Rosenblum, Z. Z. Zhanabaev, M. Kyzgarina,
and A. Pikovsky, Phys. Rev. E \textbf{80}, 046211 (2009).}

\bibitem{MN14}{O. V. Maslennikov, and V. I. Nekorkin, Phys. Rev. E \textbf{90}, 012901 (2014).}

\bibitem{MKRN13}{O. V. Maslennikov, D. V. Kasatkin, N. F. Rulkov, and V. I. Nekorkin,
Phys. Rev. E \textbf{88}, 042907 (2013).}

\bibitem{MNK15}{O. V. Maslennikov, V. I. Nekorkin, and J. Kurths,
Phys. Rev. E \textbf{92}, 042803 (2015).}

\bibitem{MN15}{O. V. Maslennikov, and V. I. Nekorkin,
Commun. Nonlinear Sci. Numer. Simul. \textbf{23}, 10 (2015).}

\bibitem{R02}{N. F. Rulkov, Phys. Rev. E \textbf{65}, 041922 (2002).}

\bibitem{RTB04}{N. F. Rulkov, I. Timofeev, and M. Bazhenov, J. Comput. Neurosci. \textbf{17}, 203 (2004).}

\bibitem{WL07}{D. Q. Wei, and X. S. Luo, Europhys. Lett. \textbf{77}, 68004 (2007).}

\bibitem{WDPC08}{Q.Y. Wang, Z. Duan, M. Perc, and G. Chen, Europhys. Lett. \textbf{83}, 50008 (2008).}

\bibitem{BPVL07}{C. A. S. Batista, A. M. Batista, J. A. C. de Pontes, R. L. Viana, and S. R. Lopes,
Phys. Rev. E \textbf{76}, 016218 (2007).}

\bibitem{ICS11}{B. Ibarz, J. M. Casado, and M. A. F. Sanju\'{a}n, Phys. Rep.
\textbf{501}, 1 (2011).}

\bibitem{FM10}{I. Franovi\'c, and V. Miljkovi\'c, Europhys. Lett. \textbf{92}, 68007 (2010).}

\bibitem{FM11}{I. Franovi\'c, and V. Miljkovi\'c, Commun. Nonlinear Sci. Numer. Simul.
\textbf{16}, 623 (2011).}

\bibitem{I06}{E. M. Izhikevich, Neural Comput. \textbf{18}, 245 (2006).}

\bibitem{IE08}{E. M. Izhikevich, and G. M. Edelman, Proc. Natl. Acad. Sci.
USA \textbf{105}, 3593 (2008).}

\bibitem{FB05}{S. E. Folias, and P. C. Bressloff, Phys. Rev. Lett. \textbf{95}, 208107 (2005).}

\bibitem{LTGE02}{C. R. Laing, W. C. Troy, B. Gutkin, and G. B. Ermentrout,
SIAM J. Appl. Math. \textbf{63}, 62 (2002).}

\bibitem{B10}{P. C. Bressloff, Phys. Rev. E \textbf{82}, 051903 (2010).}

\bibitem{BC07}{M. A. Buice, and J. D. Cowan, Phys. Rev. E \textbf{75}, 051919 (2007).}

\bibitem{BH99}{N. Brunel, and V. Hakim, Neural Comput. \textbf{11}, 1621 (1999).}

\bibitem{H03}{H. Hasegawa, Phys. Rev. E \textbf{67}, 041903 (2003).}

\bibitem{LGNS04}{B. Lindner, J. Garcia-Ojalvo, A. Neiman, and L. Schimansky-Geier,
Phys. Rep. \textbf{392}, 321 (2004).}

\bibitem{FTVB14}{I. Franovi\'c, K. Todorovi\'c, N. Vasovi\'c, and N. Buri\'c,
Phys. Rev. E \textbf{89}, 022926 (2014).}

\bibitem{FTVB13}{I. Franovi\'c, K. Todorovi\'c, N. Vasovi\'c, and N. Buri\'c,
Phys. Rev. E \textbf{87}, 012922 (2013).}

\bibitem{FTVB12}{I. Franovi\'c, K. Todorovi\'c, N. Vasovi\'c, and N. Buri\'c,
Chaos \textbf{22}, 033147 (2012).}

\bibitem{KF15}{V. Klinshov, and I. Franovi\'c, Phys. Rev. E \textbf{92}, 062813 (2015).}

\bibitem{ZSSN05}{M. A. Zaks, X. Sailer, L. Schimansky-Geier, and A. B. Neiman,
Chaos \textbf{15}, 026117 (2005).}

\bibitem{SZNS13}{B. Sonnenschein, M. A. Zaks, A. B. Neiman, and L. Schimansky-Geier,
Eur. Phys. J. Special Topics \textbf{222}, 2517 (2013).}


\bibitem{NV07}{V. I. Nekorkin, and L. V. Vdovin, Izv. Vyssh. Uchebn. Zaved.
Prikladn. Nelinejn. Din \textbf{15}, 36 (2007).}

\bibitem{CNV07}{M. Courbage, V. I. Nekorkin, and L. V. Vdovin, Chaos \textbf{17}, 043109 (2007).}

\bibitem{MN13}{O. V. Maslennikov, and V. I. Nekorkin, Chaos \textbf{23}, 023129 (2013).}

\bibitem{MN14b}{O. V. Maslennikov, and V. I. Nekorkin, "Map-Based Approach to Problems of Spiking
Neural Network Dynamics" in \emph{Nonlinear Dynamics and Complexity}, V. Afraimovich, A. C. J. Luo,
and X. Fu (Editors) (Springer International Publishing Switzerland, 2014).}

\bibitem{A99}{L. Arnold, \emph{Random Dynamical Systems}, (SpringerVerlag, Berlin,
1999).}

\bibitem{AB04}{J. A. Acebr\'{o}n, A. R. Bulsara, and W.-J. Rappel, Phys. Rev. E
\textbf{69}, 026202 (2004).}

\bibitem{GBV11}{M. Gaudreault, J. M. Berbert, and J. Vi\~{n}als, Phys. Rev. E \textbf{83},
011903 (2011).}

\bibitem{KSM10}{P. Kaluza, C. Strege, and H. Meyer-Ortmanns, Phys. Rev. E 82, 036104 (2010).}

\bibitem{TP01}{S. Tanabe and K. Pakdaman, Phys. Rev. E \textbf{63}, 031911 (2001).}

\bibitem{NM11}{V. I. Nekorkin, and O. V. Maslennikov, Radiophys. Quantum Electron.
(Engl. Transl.) \textbf{54}, 56 (2011).}

\bibitem{CMN12}{M. Courbage, O. V. Maslennikov, and V. I. Nekorkin, Chaos Soliton. Fract.
\textbf{45}, 645 (2012).}

\bibitem{MN12}{O. V. Maslennikov, and V. I. Nekorkin, Radiophys. Quantum Electron.
(Engl. Transl.) \textbf{55}, 198 (2012).}

\bibitem{G04}{C. W. Gardiner, \emph{Handbook of Stochastic Methods for Physics,
Chemistry and the Natural Sciences}, 3rd ed. (Springer-Verlag, Berlin, 2004).}

\bibitem{SPB12}{\emph{Phase Response Curves in Neuroscience: Theory, Experiment,
and Analysis}, edited by N. W. Schultheiss, A. A. Prinz, and R. J. Butera,
(Springer, New York, 2012).}

\bibitem{T07}{P. A. Tass, \emph{Phase Resetting in Medicine and Biology: Stochastic
Modeling and Data Analysis} (Springer, Berlin, Heidelberg, 2007).}

\bibitem{I07}{E. M. Izhikevich, \emph{Dynamical Systems in Neuroscience: The
Geometry of Excitability and Bursting} (MIT Press, Cambridge,
2007), Chap. 10.}

\bibitem{C06}{C. C. Canavier, Scholarpedia \textbf{1}, 1332 (2006).}

\end{thebibliography}
\end {document}